# Fueling Dynamics towards Tunable Liquid Metal Machine

Jingyi Li,[1,2] Minghui Guo,[1] Ju Wang,[1,2] Xi Zhao,[3,4,*] Jing Liu[1,2,*]

[1] *State Key Laboratory of Cryogenic Science and Technology, Technical Institute of Physics and Chemistry, Chinese Academy of Sciences, Beijing 100190, China.*

[2] *School of Future Technology, University of Chinese Academy of Sciences, Beijing 100049, China.*

[3] *Tianjin Key Laboratory for Advanced Mechatronic System Design and Intelligent Control, School of Mechanical Engineering, Tianjin University of Technology, Tianjin 300384, China.*

[4] *National Demonstration Center for Experimental Mechanical and Electrical Engineering Education (Tianjin University of Technology).*

\* E-mail: zhaoxi@mail.ipc.ac.cn, jliu@mail.ipc.ac.cn

**Abstract**

Self-propelled liquid metal-aluminum hybrid machines represent a promising class of autonomous motion systems capable of sustained movement without external power sources. While interactions between machines and their environment inevitably occur, the fundamental question of how spatial confinement affects the motion dynamics and the controllability of speed, direction, and lifetime of such liquid metal machines (LMMs) remains underexplored. Understanding these confined dynamics is essential for practical applications. Here, we present a comprehensive investigation of the non-symmetrical fueling principle governing the direction-tuning effect in LMMs. By confining LMMs within one-dimensional semi-open channels, we thoroughly disclose their impact and turning dynamics with different end obstacles throughout their lifecycle, with particular focus on fuel region morphological evolution, overall motion, and local flow characteristics after reaction times exceeding one hour. Utilizing ultra-high-speed imaging techniques, we systematically clarify how fuel region evolution and end-obstacle interactions influence symmetry-breaking mechanisms and reciprocating dynamics. Our findings reveal complex interactions between material properties, charge transfer processes, and fluid dynamics during end-turning processes, establishing a theoretical foundation for LMM driving dynamics. Beyond the theoretical mechanisms, we further demonstrate that LMM exhibits efficient heat and mass transfer capabilities, paving the way for applications in controlled transport systems and autonomous robotics.

# 1. Introduction

The creation of autonomous machines across diverse scales, dimensions, and environments has long been a human aspiration, driving extensive research. [1-4] The latest findings on the fundamentals of liquid metals (LMs) opened new avenues in robotics exploration. LMs, particularly gallium-based alloys, have emerged as highly promising functional materials in materials science and engineering. [5, 6] Their unique combination of metallic and fluid properties demonstrates significant potential in frontier fields such as soft electronics, microfluidics, and autonomous systems. [7-10] Notably, LMs exhibit remarkable autonomous behaviors and diverse responses under various stimuli. [11-13] Among these phenomena, the self-propulsion of gallium-based LM droplets fueled by aluminum (Al) stands out as particularly intriguing. [14-16] This motion is primarily governed by surface tension gradient imbalances initiated through intermetallic alloying, chemical reactions, and electrochemical processes, forming a hybrid system termed as liquid metal machine (LMM). [17-19]

Prior research has predominantly investigated the LMM dynamics within two-dimensional, weakly confined environments, such as large Petri dishes. These studies consistently reveal motion characterized by Brownian-like randomness, high disorder, and unpredictability. [20-22] Consequently, significant efforts have been directed towards constraining this chaotic motion using external fields like electric or magnetic fields. [16, 23] However, a critical gap remains in understanding how spatial confinement itself affects LMM motion dynamics and long-term operational sustainability. Only the pioneering work on LMMs in 2015 has mentioned that LMMs navigating Z-shaped tracks exhibited a pause at turns before selecting a new direction, reminiscent of the hesitation and thinking behavior in organisms navigating obstacles. [15] However, further detailed characterization of this movement and its underlying mechanisms was not provided. This research gap is especially significant given the key application scenarios envisioned for LMMs, such as vascular occlusion, targeted drug delivery, and material transport, which inherently demand predictable operation within constrained geometries and necessitate a profound understanding of LMM interactions with boundaries and obstacles during long-term operation. [24]

Furthermore, from the perspective of fundamental research in fluid dynamics, the impact and turning dynamics of LMMs upon encountering obstacles present rich and complex hydrodynamic phenomena. Unlike rigid body collisions, LMMs represent a unique fluidic system characterized by their inherent fluidity, significant deformability, and extremely high surface tension. The multiphase system encompassing LMs, aluminum fuel, and electrolyte generates intricate electrochemical interactions. [17, 25] The LMM's electric double layer modulates surface tension, while Marangoni flows and interfacial relative motions generate complex mass transport at phase boundaries. [26] What makes it even more complicated is that the physical properties and dynamic behavior of LMMs may be changed during their contact with different obstacle materials due to their inherent chemical activity. The consequences of these material-dependent interactions are still totally unknown and unpredictable. These factors collectively make it challenging to characterize the impact dynamics of LMMs.

Addressing these critical gaps, this study presents the first comprehensive investigation of the sustained reciprocating motion of millimeter-scale LMMs confined within one-dimensional semi-open channels, interacting with different constricting obstacles. Departing from prior work focused on short-term behavior, we meticulously track the entire lifecycle of these LMMs - from the initial aluminum engulfment phase, through extended active impact-turning periods, to eventual motion cessation. Leveraging ultra-high-speed high-resolution imaging techniques, we resolve the morphological evolution of the fuel region (size, morphology, orientation) and the behavior of micro-bubbles throughout the dynamic reaction and motion process. This capability enables us to systematically elucidate how local fuel consumption kinetics and end-wall interactions govern symmetry-breaking mechanisms and reciprocating dynamics during direction reversal processes, thereby revealing unprecedented local flow patterns and material restructuring mechanisms. Furthermore, we systematically clarify the influence of different confining obstacle materials (quartz, platinum, graphite) on the spontaneous impact and turning characteristics, uncovering the critical role of charge transfer and interfacial interactions in regulating reciprocating motion. This fundamental understanding enables the modulation of system behavior through strategic boundary condition engineering. Finally, we further demonstrate the practical utility of these self-fueled LMMs in enhanced heat and mass transfer applications.

These fundamental characterizations and mechanistic insights establish a theoretical foundation for LMM driving dynamics. Our findings bridge critical knowledge gaps in LMM collision dynamics and long-term autonomous behavior in confined spaces, paving the way for the transition of LMM study from previously uncontrollable irregular motion toward sustained, predictable operation. The insights and control strategies presented herein lay the foundation for functional applications in flexible robotics, chemical reactors, and adaptive thermal management systems.

## 2. Formation and Characterization of Self-propelled Liquid Metal Machines

LMMs exhibit multidirectional chaotic motion in unrestricted solution environments, particularly when interacting with container boundaries (**Figure 1a**). In this section, we introduce the formation process of LMMs and systematically characterize their motion behavior in constrained environments, with a focus on their sustained reciprocating dynamics in one-dimensional channels.

### 2.1 Aluminum-driven LM activation process

The self-propelled ability of LMMs is enabled by an activation process that typically occurs within 10 minutes (**Figure 1b**). When LM droplets contact aluminum in alkaline electrolyte solutions (**Figure 1b$_1$**), the high-surface-tension LM initially wets the aluminum surface. Subsequently, gallium rapidly penetrates and disrupts the protective $Al_2O_3$ layer, forming Al-Ga intermetallic compounds at the interface while simultaneously establishing a galvanic cell between the aluminum and the surrounding electrolyte (**Equation 1**). [27] As gallium continues to diffuse into the aluminum lattice, hydrogen bubbles form on the aluminum surface through the electrochemical reaction (**Equation 2**), causing progressive embrittlement of the aluminum structure and eventual fragmentation into discrete particles (**Figure 1b$_2$**). [28]

$$xGa(l) + (1-x)Al(s) \rightarrow Ga_xAl_{1-x}(l) \tag{1}$$

$$2Al + 2NaOH + 6H_2O \rightarrow 2NaAl(OH)_4 + 3H_2 \uparrow \tag{2}$$

Due to density differences (Al: 2.70 g·cm$^{-3}$, GaIn$_{10}$: 6.25 g cm$^{-3}$), [29] aluminum fragments sequentially float to the LM droplet surface during the penetration process (**Figure 1b$_3$**). Aluminum sheets gradually fragment and merge (**Figure 1b$_4$**), with undissolved aluminum fragments completely covering the droplet surface through liquid shear stress and surface tension effects (**Figure 1b$_5$**). As aluminum fragments dissolve and oxidation reactions consume material, the coverage area gradually decreases (**Figure 1b$_6$**). Spatially heterogeneous aluminum fragment coverage induces chemical potential gradients across the droplet surface, generating Marangoni convection.[30] Directional migration toward regions of lower aluminum fragment density occurs when the resulting interfacial tension gradient exceeds the threshold required to overcome viscous drag forces (**Figure 1b$_7$**). Progressive fragment dissolution reduces the aluminum content to a residual particle (**Figure 1b$_8$-1b$_9$**), which converges at the droplet-substrate interface through synergistic hydrodynamic mechanisms, establishing a stable propulsive configuration that marks the transition to sustained autonomous locomotion. We define the solid fragment aggregation area responsible for hydrogen bubble release on the surface of the resulting LMM as the "fuel region". This domain generates surface tension gradients and hydrogen bubble release through localized electrochemical processes, providing the driving forces for sustained autonomous motion.

Once activated, LMMs exhibit highly random motion in two-dimensional planes, undergoing multiple collisions with container boundaries (**Figure 1c**). However, the stochastic nature of their movement makes collision points unpredictable, making it difficult to strategically position measurement points for capturing and analyzing velocity and morphological changes during collision processes.

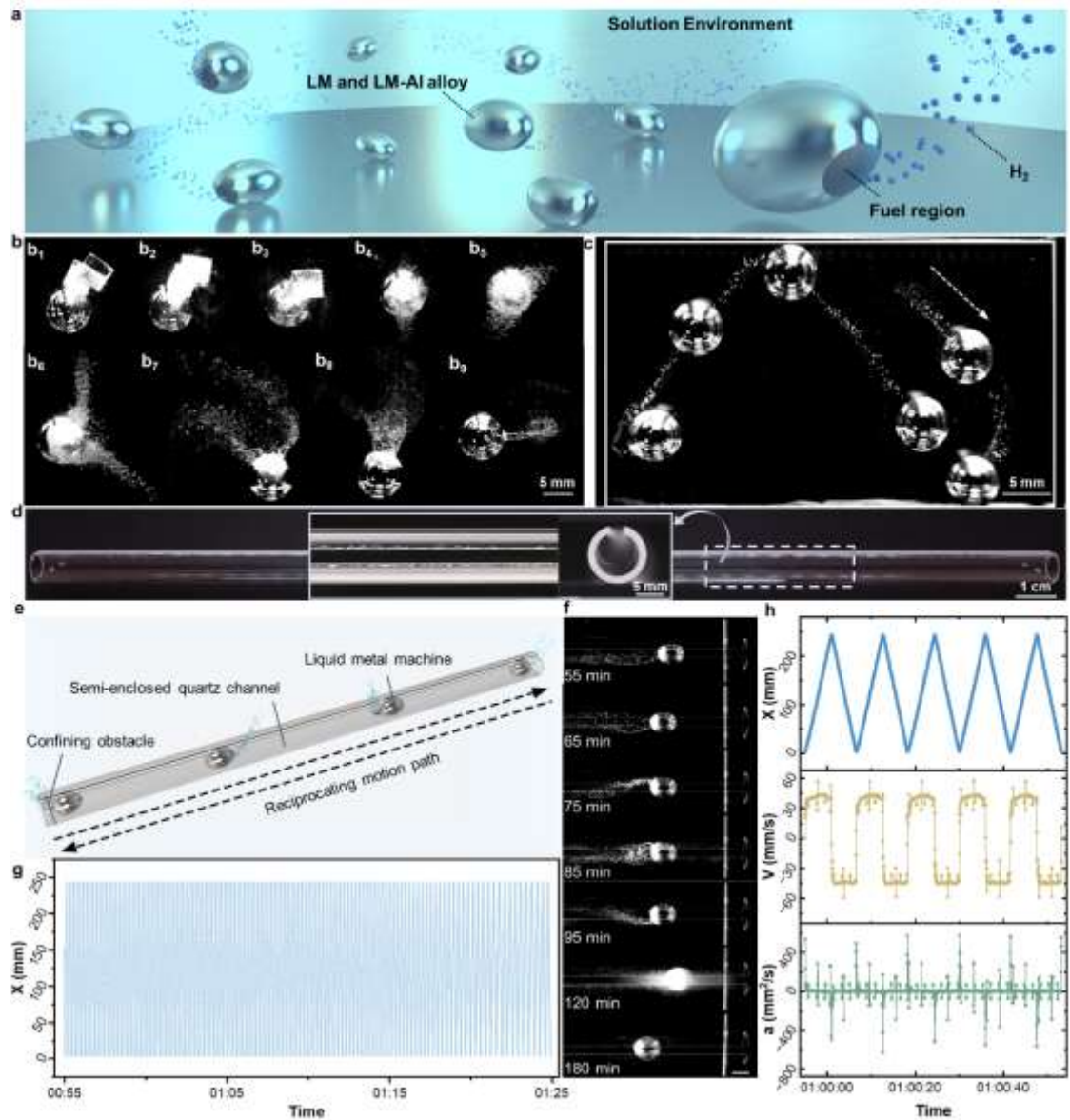

**Figure 1. Formation and characteristics of LMMs.** a. LMMs demonstrating multidirectional motion and wall collisions in free space solution environments; b. Process of LMM formation; c. Self-propelled LMMs undergoing multiple impacts and rebounds in a two-dimensional confined space; d. Quartz channel designed to restrict LMMs' motion to one dimension; e. LMMs performing directional motion within the channel, returning after end impact to conduct sustained reciprocating motion; f. Temporal morphological evolution of LMMs within the channel (scale bar: 4 mm); g. Displacement variations of the LMM during 55-85 minutes following initiation of reaction; h. Relationships between displacement, velocity, and acceleration of the LMM during motion.

## 2.2 One-dimensional confinement and sustained reciprocating motion

To enable systematic study of collision dynamics, we confine LMMs within one-dimensional channels. We designed a quartz channel (inner diameter 6 mm) featuring rod-shaped confining obstacles (diameter 0.7 mm) at both ends and a top opening design that prevents bubble accumulation during motion (**Figure 1d**). This configuration restricts LMM motion to a single axis while preserving the chemical environment required for operation, enabling precise study of collision and directional reversal mechanisms in confined spaces. When placed in this channel, LMMs exhibit pronounced spontaneous reciprocating one-dimensional motion (**Figure 1e**). They move along the channel to one end, collide with the quartz obstacles crossing the central axis, then reverse direction and continue toward the opposite end. Upon reaching another end, directional reversal occurs again, forming a sustained reciprocating motion lasting approximately 2 hours.

The morphology of LMMs and their bubble trails continuously evolves throughout their operational lifetime (**Figure 1f**). Initially, the LMM maintains a smooth surface with the fuel region distributed at its tail. As reciprocating motion continues, we observe progressive changes in the tail fuel region area and bubble release density. By 2 hours, the LMM surface becomes completely covered with fuel region, generating dense bubbles throughout before motion ceases. At 3 hours, the LMM surface recovers its metallic luster and pure LM composition. These changes reflect ongoing electrochemical reactions, fuel consumption, and interactions with channel walls. All these factors influence propulsion mechanisms and motion characteristics over the extended time scales examined in this paper.

To quantitatively characterize the long-term behavior of this system, we tracked LMM displacement during an extended period (55-85 minutes) after reaction initiation. The data reveals a sustained reciprocating motion pattern with remarkable temporal consistency (**Figure 1g**). Through detailed analysis of five complete cycles, we systematically examined relationships between LMM displacement, velocity, and acceleration (**Figure 1h**). Results indicate that the LMM maintains relatively uniform motion in the central channel region while experiencing distinct acceleration and deceleration phases within short distances near the ends.

These preliminary observations raise fundamental scientific questions: What physicochemical changes occur during end collisions in the spontaneous reciprocating process of LMMs? How does the system's long-term dynamic behavior evolve over time? Which factors can regulate their motion characteristics? The following sections address these questions, exploring mechanisms controlling reciprocating behavior, focusing on symmetry-breaking phenomena during dual-end collisions, spatiotemporal evolution characteristics of reciprocating patterns, and the regulation effects of confining obstacle material properties on overall system dynamics.

## 3. Symmetry Breaking Mechanisms and Directional Reversal Dynamics

The reciprocating motion of LMMs essentially depends on non-equilibrium state maintenance mechanisms that enable directional propulsion and collision-turning processes (**Figure 2a**). This section investigates the evolution of these mechanisms during the system's operational lifetime, with particular focus on the influence of fuel region dynamics.

**3.1 Fuel region distribution and its impact on directional propulsion**

Fuel regions exhibit distinct spatial distributions that directly correlate with velocity magnitude and directional stability, and undergo, systematic evolution over motion time (**Figure 2b**). Initially, fuel regions occupy only approximately 1% of the visible LMM surface area and remain relatively stable. Their area rapidly expands after 70 minutes, accompanied by reduced motion velocity that causes changes in bubble discharge pathways and bubble accumulation at the solution surface (**Figure 2c**). Fuel regions consistently concentrate in the trailing region of the advancing droplet due to multiple hydrodynamic effects, including ambient fluid shear forces, Marangoni-induced surface circulation, and tail vortex entrainment mechanisms. [31] Bubble-generating regions consistently remain at the upper edge of fuel regions because the trailing coverage does not consist entirely of aluminum fragments as reactions proceed. The bottom portion of the fuel region accumulates reaction products, including intermetallic phases and aluminum hydroxide precipitates. [32] Concurrently, interfacial energy variations at the liquid-solid-gas triple-phase boundary promote continuous wetting and spreading of the coverage across the surface. As the fuel region area increases, the active aluminum fragment distribution progressively migrates toward the droplet apex, eventually achieving uniform surface coverage. This transition eliminates the spatial asymmetry in surface chemical potential and reaction site distribution, reducing surface tension gradients below the threshold required to overcome viscous dissipation and maintain self-propulsive motion.

As the area and composition of this fuel region evolve, the symmetry-breaking mechanism of LMMs undergoes significant changes. At approximately 75 minutes, the initially enlarged fuel region at the tail of directionally advancing LMMs leads to a distinctive "∞" motion trajectory (**Figure 2d**). This phenomenon results from the combined effects of thrust point upward migration due to preliminary accumulation of reaction products at the lower end of fuel regions, and the persistent presence of dual symmetric vortices behind the droplet. We designate this temporal threshold as the critical boundary distinguishing early and late stages of LMM motion, where end impact-turning process exhibits distinct stage-dependent characteristics. Notably, LMMs operating in unrestricted two-dimensional space also exhibit similar ∞-shaped motion trajectories, confirming this behavioral signature's independence from spatial constraints (**Figure 2e**).

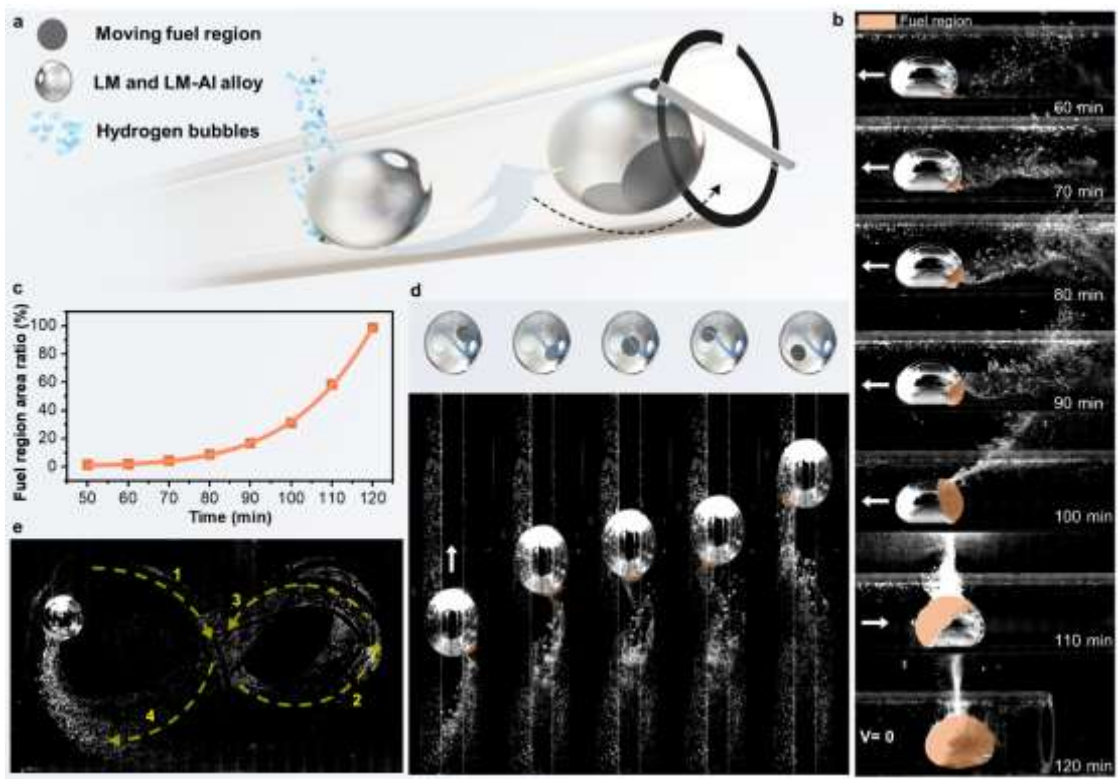

**Figure 2. Impact of fuel region distribution on LMM directional propulsion.** a. LMM uniform motion and end turning processes in one-dimensional channels, accompanied by fuel region changes; b. LMM morphological and bubble trajectory evolution during 1-2 hours of continuous operation (fuel regions highlighted, LMM volume $V_{LMM}$: 100 μL), white trails at the tail represent hydrogen bubbles; c. Changes in fuel region distribution area and the inverse relationship between motion direction and fuel region distribution during 50-120 min of continuous operation; d. Indicator of entering late-stage motion: fuel regions at LMM tail exhibit "∞" path movement during overall forward motion ($V_{LMM}$= 100 μL), with fuel regions highlighted; e. In a free two-dimensional environment, LMM's overall movement path presents the "∞" shape ($V_{LMM}$= 100 μL).

### 3.2 End impact-turning mechanism

The dynamic characteristics of LMMs during directional transitions at channel ends are significantly influenced by fuel region distribution and exhibit distinct temporal patterns. Before contacting the quartz obstacles at the channel ends, LMMs maintain approximately uniform motion through dynamic equilibrium between two opposing force groups (**Figure 3a$_1$, 3b$_1$, and 3c$_1$**): resistive forces including viscous drag from surrounding fluid ($F_v$) and frictional resistance from channel surfaces ($F_f$), and propulsive forces including surface tension gradient-driven propulsion from electrochemical reactions ($F_e$) and reactive thrust from hydrogen bubble release ($F_H$). Upon contact with the end obstacles, LMMs enter the deceleration-deformation stage. Due to their inherent flow characteristics, LMMs undergo significant deformation while maintaining their original motion direction (**Figure 3a₂** and **3c₂**), but with gradually decreasing velocity. The mechanical system becomes more complex, incorporating lateral support forces ($F_s$) from the obstacles and capillary restoring forces ($F_c$) arising from surface tension acting on the deformed interface in addition to the original propulsive and resistive forces (**Figure 3b₂**). Notably, despite good fluidity, the high surface tension of LMs limits deformation extent, preventing passage through the obstacles positioned at half the channel height.

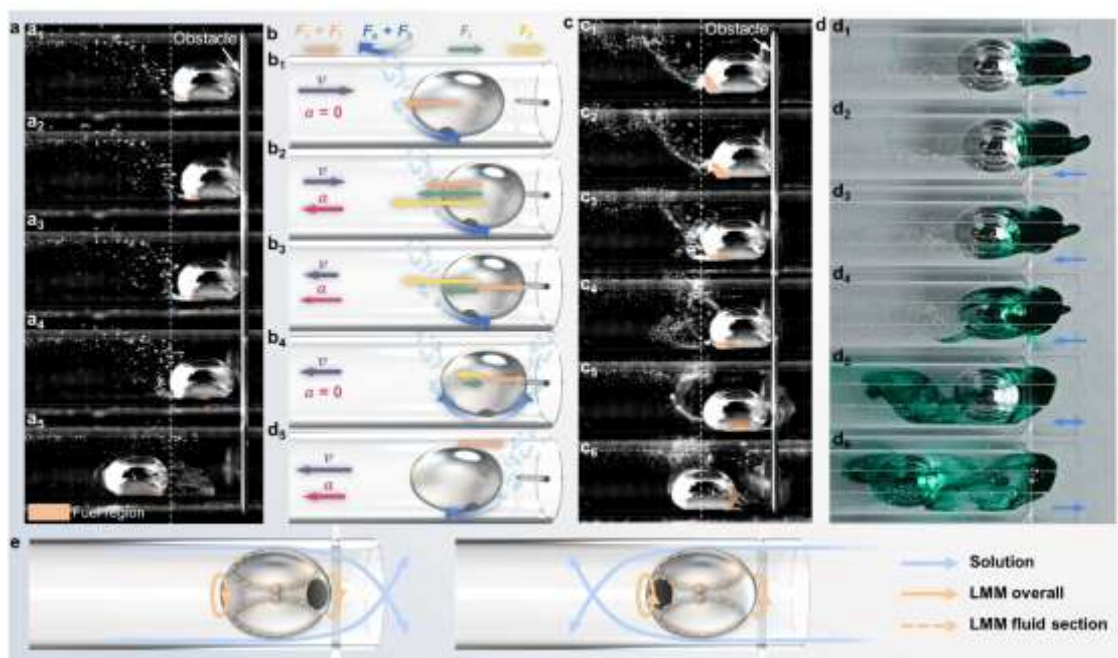

**Figure 3. Impact of fuel region distribution on LMM impact-turning processes.** a. Side view of LMM during the right end impact-turning process at 1 hour of continuous operation in the early stage (fuel regions highlighted, $V_{LMM}$= 100 μL): a₁. Instant of LMM contact with obstacle; a₂. LMM moving towards maximum deformation; a₃. LMM rebounds with the fuel region maintaining its absolute spatial coordinates while the fluid section translates in the opposite direction; a₄. LMM fluid portion continues to move leftward while fuel region position transfers to the bottom of the overall body; a₅. Fuel region transfers to the right tail and propels LMM to accelerate forward. b. Detailed Force analysis of LMM during end impact-turning process and the corresponding velocity ($v$) and acceleration ($a$) evolution. Force vectors include: propulsive forces ($F_e$: electrochemical propulsion, $F_h$: hydrogen bubble thrust), resistive forces ($F_v$: viscous drag, $F_f$: frictional resistance), and collision-induced forces ($F_s$: lateral support force, $F_c$: capillary restoring force). Arrow lengths indicate relative force magnitudes, and directions show force orientations relative to the instantaneous motion. b₁. Uniform motion phase; b₂. Deceleration-deformation phase: additional forces $F_s$ and $F_c$ emerge upon obstacle contact; b₃. Shape recovery phase: $F_e$ temporarily opposes the new motion direction as fuel region maintains its absolute position due to inertial effects and differential adhesion; b₄. LMM reaches Transitional state with fuel region relocated to bottom position, where horizontal component of $F_e$

vanishes; b₅. Accelerated motion: fuel region transfers to new tail position, $F_e$ realigns with reversed motion direction; c. Side views of LMM achieving the directional change process at 80 min of continuous operation in the late stage (fuel regions highlighted, $V_{LMM}$= 100 μL), showing multi-impact oscillatory behavior due to reduced kinetic energy and enhanced fuel region adhesion; d. Dye-traced flow of the surrounding fluid during the LMM impact-turning process in the late stage ($V_{LMM}$= 100 μL), revealing flow reversal patterns correlated with fuel region positioning; e. Motion paths of external fluid, internal fluid, and overall of the LMM before and after LMM directional change.

After deformation reaches its maximum, LMMs enter the shape recovery stage (**Figure 3a₃** and **3c₃**). However, due to inertial effects and differences in adhesion between different parts and the channel, the fuel region maintains its absolute spatial position, while the generated propulsive force temporarily opposes the new motion direction, causing continued LMM deceleration (**Figure 3b₃**). As the fluid portion recovers its shape and continues moving leftward until momentary rest (**Figure 3b₄**), the fuel region relocates to the lower portion of the LMM (**Figure 3a₄**), after which fuel region behavior differs between early and late stages. In the early stage, LMM fuel regions with small areas and relatively simple composition smoothly transfer to the right-side bottom of the LMM and drive leftward acceleration until propulsive and resistive forces reestablish equilibrium (**Figure 3a₅**). However, in the later stage, due to factors including reduced motion velocity, increased fuel region area, and enhanced adhesion of aluminum hydroxide depositing in the lower section, portions of the fuel region remain at the tail. This causes the action point of driving forces generated from the upper effective reaction zone to shift downward, increasing the horizontal component of propulsive forces and causing the LMM to move toward the end again (**Figure 3c₄**). As motion time increases, this short-distance repetitive cycle of impact, deformation, and rebound occurs more frequently. Multiple oscillations lead to momentum dissipation and loosening of fuel region adhesion to the bottom (**Figure 3c₅**), ultimately enabling the fuel region transfer to the right-side bottom and reestablishment of the driving mechanism (**Figure 3c₆**).

During LMM motion, three components of the entire system continuously interact and influence each other: the internal fluid portion of the LMM, the fuel region, and the external solution. To further investigate the variation patterns of these fluid components and elucidate how the fuel region influences the symmetry-breaking mechanisms of LMMs, we employ fluid tracing methods to track the flow trajectories of the external solution. Before fuel region migration to the bottom position, external fluid is driven by internal fluid motion and flows from outside the channel inward along the LMM-wall gap (**Figure 3d₁-d₄**). During multiple impact-turning cycles, this flow experiences perturbations and instabilities (**Figure 3d₅**). Upon completion of the fuel region positional transition, external fluid flow reverses direction, moving from inside the channel outward (**Figure 3d₆**).

Furthermore, tracer streamlines reveal three-dimensional helical flow patterns rather than simple unidirectional flow, indicating that the LMM surface not only exhibits dual toroidal flow fields but also undergoes overall rotational motion (**Figure 3e**). [33] The coupling between internal fluid circulation and external flow patterns creates a hydrodynamic system where surface tension gradients and momentum transfer collectively drive the propulsion mechanism. These results demonstrate that fuel region positioning directly controls the external fluid flow patterns, establishing a clear mechanistic link between fuel region evolution and propulsion dynamics.

### 4. Tuning Reciprocating Motion Through Interface Engineering

Our analysis of LMM behavior in quartz obstacle systems reveals that LMM propulsion and turning dynamics are fundamentally governed by fuel region evolution. This observation suggests that interfacial interactions at obstacle contacts may provide a controllable mechanism for modulating LMM behavior. LMs process inherent chemical activity that can generate diverse electrochemical responses upon contact with different materials, potentially altering both surface charge distribution and interfacial energy landscapes. [34] To explore this hypothesis, this section systematically investigates how obstacle material properties influence the symmetry-breaking mechanisms through direct interfacial contact.

We employ three representative obstacle materials at both channel ends: quartz (non-metallic insulator), platinum (metallic conductor), and graphite (non-metallic conductor), maintaining identical dimensions across all samples (**Figure 4a**). This material selection spans the spectrum from purely mechanical interactions to various electrochemical coupling modes, enabling systematic investigation of how charge transfer and surface chemistry collectively regulate LMM dynamics.

#### 4.1 Effects of dual-end obstacle materials on motion characteristics

Under the influence of different material obstacles, the one-dimensional reciprocating motion of LMMs exhibits distinct differences in motion characteristic evolution, including period, velocity distribution, and intra-period time allocation. We define one period as the time interval between consecutive passages of LMMs through the channel midpoint in the same direction, with each period comprising approximately uniform directional motion and two end turning processes.

The three material systems exhibit distinct evolutionary patterns over a representative 30-minute observation window encompassing both early and late motion stages, LMMs operating in channels with quartz obstacles demonstrated the most consistent oscillation periods (**Figure 4b**). In contrast, platinum obstacle channels initially show slightly higher but relatively stable periods, which increase over time and exhibit stepwise jumps in the late stage. Graphite obstacles produce the longest periods, accompanied by significant fluctuations. Correspondingly, velocity distribution during directional motion phases shows material-dependent characteristics (**Figure 4c**). In the early stage of motion, all three materials maintain relatively stable velocities, with quartz consistently reaching the highest values, platinum showing intermediate levels, and graphite exhibiting the lowest velocities. As the motion progresses into the later stage, all systems experience a decline in velocity, but with noticeably different patterns. LMMs in quartz obstacles sustain the highest velocities throughout the entire period despite a gradual decrease. Platinum systems initially preserve intermediate velocity levels but then experience rapid fluctuations after 80 min, eventually falling below graphite levels. Graphite systems, while starting with the lowest velocities, decline more gradually compared to the dramatic changes observed in platinum systems.

We further compare the percentage of time allocated to directional motion versus dual-end turning processes within each period to reveal the differential effects of end obstacle materials on the turning process (**Figure 4d**). In quartz obstacle channels, LMMs require approximately 2.9% of the period time for dual-end turning, with this proportion gradually increasing after 80 minutes. Platinum obstacle channels require only 2.7% of their period time with a gradual decreasing trend. In contrast, graphite obstacle channels require up to 30% of the period time for dual-end turning, exhibiting significant early-stage fluctuations followed by stable, decreasing values in the late stage. These pronounced temporal distribution differences, combined with the distinct adhesion and electrochemical properties between LMMs and the three materials, reflect the capacity of end obstacles to modulate LMM motion. [29, 35]

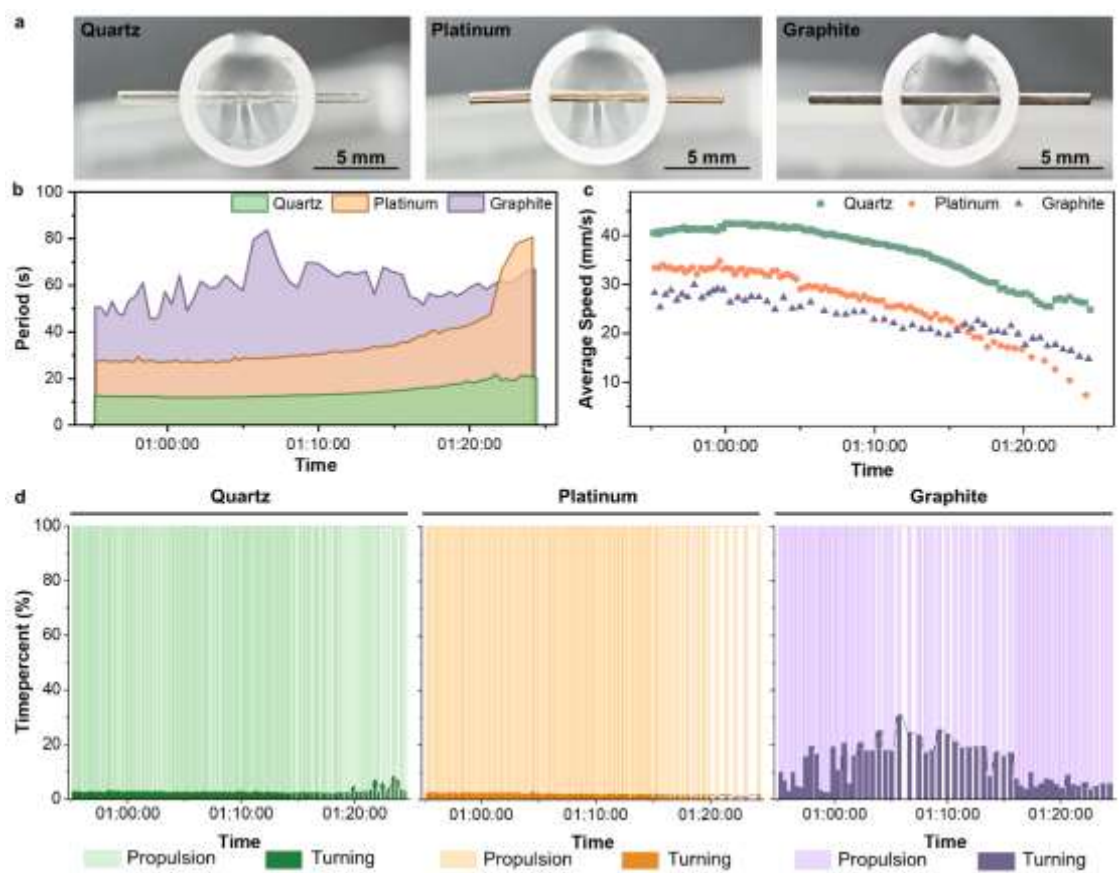

**Figure 4. Reciprocating motion characteristics of LMMs in different materials obstacle systems.** a. Three representative materials obstacles; b. Temporal evolution of motion period; c. Velocity variations during operation; d. Evolution of temporal allocation between directional motion and dual-end turning process per cycle for quartz, platinum, and graphite obstacles.

**4.2 Interfacial mechanisms underlying material-dependent motion characteristics**

As conductive materials capable of electrochemical interactions with LM/Al, platinum and graphite can significantly influence the charge imbalance state on LMM surfaces. [36] To investigate this phenomenon, we develop a monitoring system that directly measures potential variations across obstacles during LMM reciprocating motion (**Figure 5a**). This apparatus enables correlation between LMM displacement and real-time electrical measurements while recording periodic changes. We retain quartz from our previous fluid behavior studies as one obstacle material due to its insulating properties, which provide purely mechanical interactions with LMMs, thereby serving as an ideal control to distinguish electrochemical from purely physical effects.

Electrical potential measurements reveal distinct material-dependent patterns that evolve with LMM operational period (**Figure 5b**). Specifically, platinum obstacle systems display unique electrochemical characteristics during LMM contact with measurement electrodes. At approximately 55 minutes of operation, the system reveals alternating potential peaks: an initial negative peak followed by a positive peak, with rapid return to baseline after motion completion. These alternating potential peaks reflect rapid switching between two galvanic cell configurations. The initial contact establishes a Ga/In-Pt cell, which transitions to an Al-Pt cell when the fuel region interacts with platinum during turning. This transition occurs due to the more negative standard potential of $Al^{3+}/Al$ at -1.66 V, which provides a stronger electrochemical driving force compared to $Ga^{3+}/Ga$ at -0.53 V. [36-38] Platinum's high catalytic activity for hydrogen evolution accelerates fuel consumption and generates significant surface tension gradients near the fuel region. [39, 40] This enhanced electrochemical activity explains the consistently lower velocities observed in platinum systems compared to quartz systems despite similar propulsion mechanisms.

Potential alternations between positive and negative values persist throughout the motion, while peak magnitudes show a decreasing trend and after 80 min decrease by two orders of magnitude, and the duration of individual potential fluctuations significantly extends. This temporal evolution reflects two concurrent processes: diminished aluminum activity reducing electron transfer rates, and enhanced capacitive effects from the expanded fuel region area prolonging charge redistribution. The decreased catalytic capability of platinum electrodes in later stages, caused by hydroxide ion adsorption blocking active sites, contributes to the rapid transitions in motion characteristics observed after 80 minutes. [41] This temporal evolution correlates directly with the prolonged end turning times observed in kinematic analysis, while multiple directional changes in potential align with the multiple fuel region relocations characteristic of late-stage dynamics.

In contrast, graphite obstacle systems exhibit more pronounced fluctuations in potential change duration within each cycle, with LMM-obstacle contact producing potential fluctuations tens of times higher than those in platinum systems (**Figure 5c**). The substantially higher initial potential peaks arise from graphite's significantly lower electrocatalytic activity for hydrogen evolution, requiring higher overpotentials to drive comparable reaction rates. [42-44] Additionally, the hierarchical porous microstructure of graphite provides large effective surface area, enhancing charge separation and double-layer capacitance effects. [44, 45] The complex adsorption-desorption processes on graphite surfaces create additional energy dissipation pathways beyond simple electrochemical reactions. Energy conversion in graphite systems follows a multistep sequence where chemical energy first transforms into electrochemical energy, then converts sequentially to adsorption energy, surface energy, and ultimately kinetic energy, with partial dissipation as thermal energy through interlayer sliding. This complex energy coupling explains the extended turning times and significant period fluctuations characteristic of graphite systems. Similar to platinum systems, graphite systems also exhibit potential amplitude reduction with operational time, as hydrogen evolution progressively alters surface morphology and introduces defect sites that diminish electrochemical influence.

To understand the interaction mechanism differences across obstacle materials, we conduct surface analysis of end obstacles after LMM operation. Surface characterization via scanning electron microscopy (SEM) and energy-dispersive X-ray spectroscopy (EDS) elemental mapping revealed distinct morphological differences between the electrode materials. Platinum wire electrodes maintained relatively smooth surface morphologies (**Figure 5d**), whereas graphite rod electrodes exhibited characteristic hierarchical porous structures with localized Ga-In alloy accumulation and aluminum compound deposition (**Figure 5e**). These material transfer patterns directly explain the significant influence of obstacle surface properties on LMM spontaneous reciprocating motion behavior, as both wetting and electrochemical effects occur during each contact.

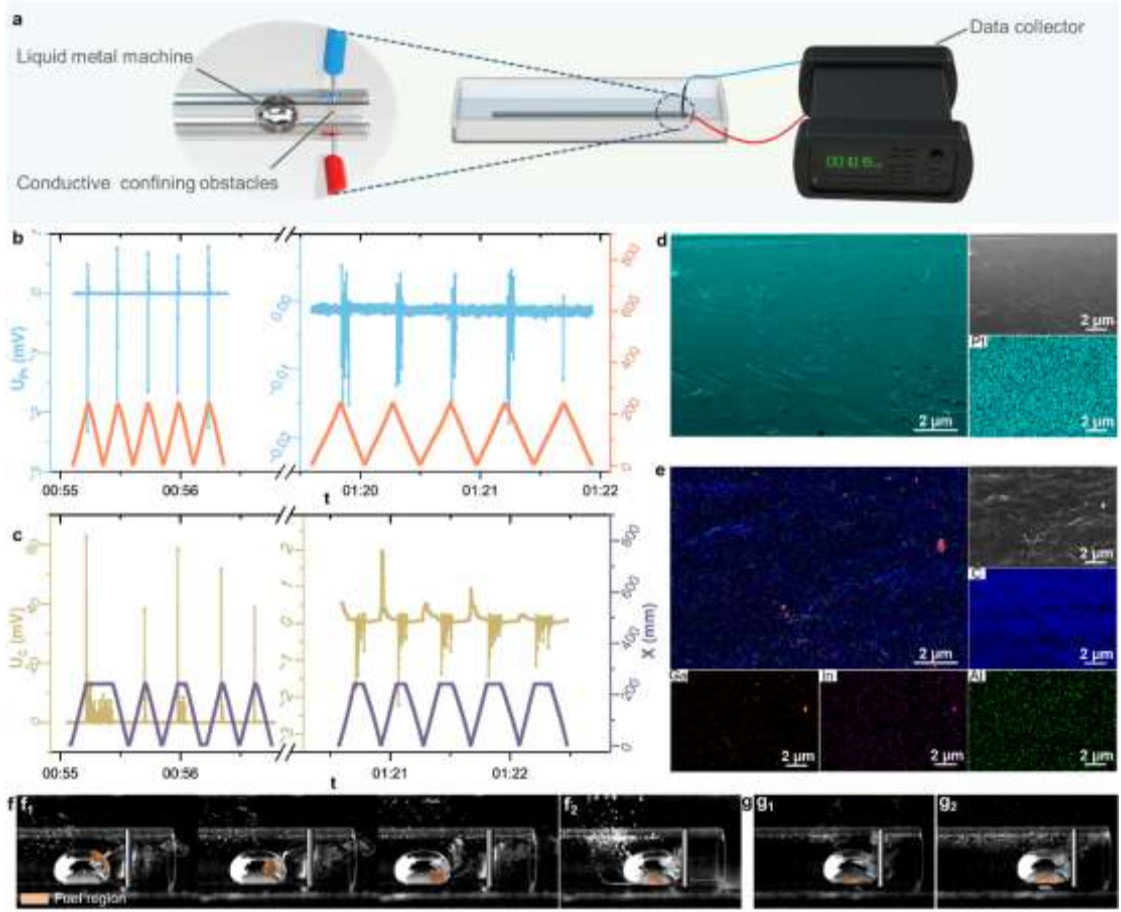

**Figure 5. End charge transfer monitoring and mechanisms of LMMs in one-dimensional systems with platinum and graphite obstacles.** a. Experimental setup for monitoring charge transfer processes during LMM end interactions; b-c. Relationship between end voltage variations and displacement distance for platinum and graphite obstacles during early and late motion stages; d-e. Post-motion SEM images and corresponding EDS elemental distribution maps of platinum wire and graphite rod used as confining obstacles; f. LMM impact-turning process with platinum obstacles (fuel regions highlighted, $V_{LMM}$= 100 μL): $f_1$. early stage, $f_2$. late stage; g. LMM impact-turning process with graphite obstacles (fuel regions highlighted, $V_{LMM}$= 100 μL): $g_1$. early stage, $g_2$. late stage.

Fuel region behavior during turning also exhibits material-dependent characteristics. Fuel regions in platinum obstacle systems do not migrate along the bottom wall to the opposite side during directional reversal as observed in quartz or graphite obstacle systems. Instead, regardless of early or late motion stages, they slide along the LMM upper surface toward the opposite end (**Figure 5f**). Moreover, due to surface tension effects, fuel regions undergo fragmentation and reorganization while sliding along the LMM upper surface. In this process, the fuel regions directly contact the platinum wire, rapidly transforming the electrode configuration from a Ga/In-Pt to an Al-Pt battery. Conversely, in graphite obstacle systems, fuel regions consistently remain distant from the electrodes when LMMs contact the obstacles, maintaining constant surface charge flow direction (**Figure 5g**).

Unlike the straightforward chemical-to-surface-to-kinetic energy conversion observed in quartz obstacle systems, where chemical energy converts through surface energy to kinetic energy, energy conversion in platinum and graphite obstacle systems exhibits greater complexity and temporal variability. In platinum systems, the high catalytic activity of platinum facilitates charge transfer processes, initially accelerating electrochemical reaction rates. [37, 46] However, when catalytic activity progressively decreases below critical thresholds, it results in rapid transitions in reciprocating motion characteristics. [38, 42] For graphite interfaces, the combination of enhanced capacitive effects, increased adsorption sites, and mechanical energy dissipation through the porous architecture collectively results in more complex dynamic characteristics. [44] Consequently, the turning time ratio within each cycle exhibits a decreasing trend over time, with late-stage end turning processes often becoming smoother than early-stage ones.

## 5. Functional Applications

The reciprocating motion behavior of LMMs enables practical applications in heat and mass transfer systems. This section demonstrates applications that leverage autonomous motion for efficient transport processes without external energy input.

### 5.1 Enhanced heat transfer capabilities

Heat transfer in closed or small-opening channel systems typically faces efficiency challenges.[47, 48] Traditional approaches primarily rely on long-distance passive conduction or externally-driven pumping mechanisms, with the former exhibiting low thermal transfer efficiency, while the latter increased system complexity and energy consumption. Although LMs can be wirelessly controlled within enclosed spaces through various external fields, [49, 50] self-propelled LMMs offer a more flexible field-free solution for heat transfer in long-distance semi-closed systems with limited energy resources. This approach combines the high thermal conductivity of LMs with autonomous motion capabilities, improving axial heat transfer efficiency within pipelines. Additionally, the passive motion of the surrounding fluid induced by LMM operation facilitates improved heat exchange between internal and external solution environments, thereby achieving multidimensional heat transfer enhancement beyond conventional static systems.[51]

To quantify this enhancement, we employ temperature-adjustable stainless-steel blocks as differential temperature sources, which do not introduce additional chemical effects on the solution environment, including influences from solute composition and concentration changes (**Figure 6a**). Thermal imaging results show significant temperature redistribution in both the channel and surrounding solution when comparing systems with LMM (1h) operation to passive heat transfer conditions (**Figure 6b**). The temperature differential is established using 0 °C cold and 50 °C hot sources positioned at opposite ends. While static systems rely primarily on conduction through channel fluid and walls for heat transfer, LMMs' turning motion between cold and hot ends drives surrounding solution flow, introducing convective heat transfer modes. This active transport generates dynamic temperature patterns that evolve with each oscillation cycle, increasing heat transport rates compared to diffusion-limited systems and reducing the time required to achieve temperature uniformity across the three regions.

We define three key measurement points along the channel centerline to analyze specific temperature changes at the cold end, center, and hot end during heat transfer processes (**Figure 6c**). During system operation, temperatures within the channel at the cold end cross-section are significantly lower than those in the external solution at the same cross-section. The cold source influence diminishes progressively with increasing distance from the measurement point. The hot end cross-section exhibits similar temperature gradient profiles, with the lowest temperatures inside the channel due to the remote cold source effects transported by LMM motion.

The thermal effects of LMM bidirectional motion are most pronounced within the channel, with the center point experiencing the largest temperature fluctuation amplitude across each cross-section. Time series analysis of temperature changes at these three key points over a 190 s window showed the dynamic characteristics of this system's heat transfer process (**Figure 6d**). Temperature curves display periodic fluctuations that closely match LMM displacement frequencies. Each LMM movement from hot to cold end causes periodic temperature increases at the cold end due to transported thermal energy. Correspondingly, the hot end exhibits periodic cooling effects as heat is actively removed. As LMMs departed, both cold and hot end temperatures gradually recovered to levels approaching their respective equilibrium zone temperatures. This thermal cycle is typically initiated with rapid temperature change peaks followed by relatively slower thermal conduction-driven recovery.

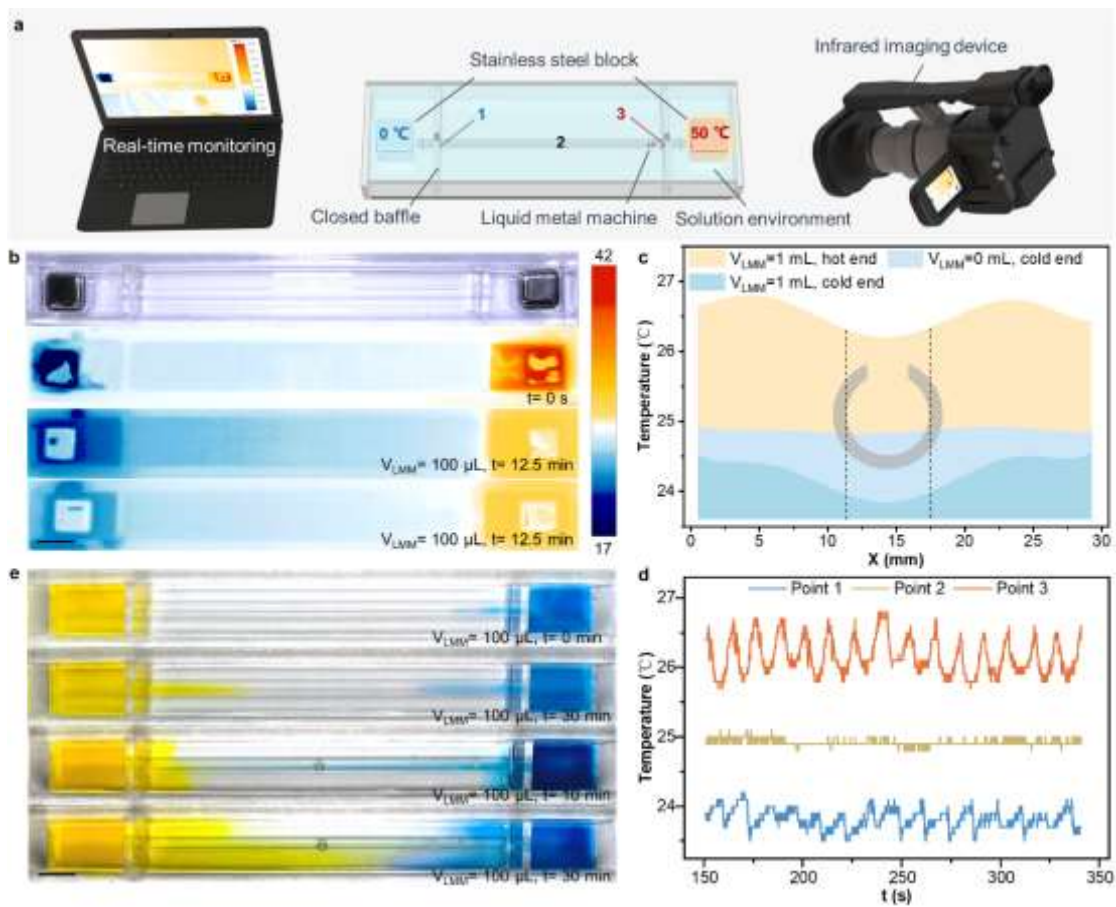

**Figure 6. Applications of LMMs in heat and mass transfer processes.** a. Experimental system for evaluating heat transfer performance of LMMs; b. Heat transfer effects in a one-dimensional motion system with 0 °C cold source and 50 °C hot source positioned at left and right ends, respectively (unit: °C, scale bar: 25 mm); c. Temperature distribution profiles at cold end, center, and hot end cross-sections at 225.328 s after heat source placement; d. Temperature evolution curves at cold end, center, and hot end during 150-340 s following heat source introduction; e. Mass transfer effects in a one-dimensional motion system with yellow and blue solutions at the left and right end regions, respectively (scale bar: 25 mm).

**5.2 Mass transport and mixing applications**

Beyond heat transfer, the reciprocating motion of LMMs can similarly enhance mass transfer and mixing in semi-closed confined spaces. To demonstrate this capability, we positioned closed solution pools of different colors at both channel ends, with the one-dimensional channel serving as the sole pathway connecting the two regions. Under static conditions, these solutions undergo slow diffusion only within the channel. Introduction of LMMs significantly accelerates this diffusion process. Each oscillation cycle transports portions of one solution toward the opposite end, forming intermediate mixing zones that spread within the central solution pool. Additional fluid displacement generated during LMM end turning further enhances local solution mixing near channel ends. Time-lapse imaging results show that the diffusion effects achieved with LMMs after 10 minutes significantly exceed those of 30-minute static diffusion (**Figure 6e**). Initially, sharp color boundaries become increasingly blurred with increasing LMM reciprocating cycles. This creates color gradients along the channel that indicate effective mass transport between ends. At ends specifically, LMMs create local high-shear and turbulent regions that significantly enhance material transport rates. [52] The millimeter-scale of the LMM dimensions makes them suitable for integration into compact fluid devices, avoiding the energy consumption associated with traditional pumping mechanisms. [53]

## 6. Conclusion

This study systematically investigates the long-term reciprocating dynamics of self-fueled liquid metal machines (LMMs) in one-dimensional semi-open channels. We reveal that sustained autonomous oriented motion and end turning are enabled by dual symmetry-breaking mechanisms: fuel region distribution and boundary interactions. The core driver of LMM motion is the asymmetric distribution of matter and charge within the fuel regions, which manifests in two distinct motion stages: an early stage characterized by reverse acceleration with fuel region sliding to the tail position after collision, and a late stage featuring complex reversals with multiple collisions and dynamic pauses where fuel regions remain at the bottom. Furthermore, obstacle material properties significantly influence LMM dynamics: Quartz (electrical insulator, purely mechanical interaction) yields stable oscillation periods and high velocities; Platinum (catalytic/electrochemical properties) extends periods and alters velocities; Graphite (wetting/electrochemical effects) induces highly fluctuating periods due to complex charge transfer and surface adhesion. Finally, we demonstrate the practical application of LMMs by showcasing their ability to enhance heat and mass transfer efficiency, leveraging their high thermal conductivity and fluid mixing during motion. This research establishes critical relationships between fuel region dynamics, obstacle properties, and motion behavior during long-term one-dimensional LMM operation, providing a theoretical foundation for directing autonomous machines in confined geometries and harnessing inherent interfacial asymmetries in soft actuators.

## 7. Experimental Section/Methods

**Synthesis of LMMs**

$GaIn_{10}$ was prepared by mixing 90 wt% gallium and 10 wt% indium, and melting the mixture in a vacuum oven at 150 ℃ for 6 hours. Aluminum foil with 99.99% purity and 0.1 mm thickness was used. $GaIn_{10}$ and aluminum were mixed at a mass ratio of 100:1.08 and placed in 0.5 mol/L NaOH solution.

**Synthesis of One-Dimensional Semi-Open Channels**

Quartz one-dimensional channels were fabricated with the following specifications: length of 250 mm, inner diameter of 6 mm, and outer diameter of 8 mm. Laser drilling was performed to create opposing holes with a diameter of 7.2 mm, positioned 5 mm from each end and 3 mm from the bottom of the inner wall. A 3 mm slot was cut parallel to the axis at the top of each channel.

The main instruments were as follows: (1) High-speed camera, used for high frame rate imaging for capturing the movement of LMMs. Model: 28028 VEO-710L-18GB-M, Phantom, America. (2) Data acquisition system, used for monitoring of end-point potential changes. Model: DAQ970A, Keysight, America. (3) Electronic balance, used for weighing NaOH and metals. Model: ZN-C50002, Hangzhou Youheng Weighing Equipment Co., Ltd. Resolution: 0.01 g. Maximum capacity: 5000 g. (4) Intelligent heating platform, used for heating LMs. Model: GJR-2020, Hangzhou Gongjiangren Technology Co., Ltd. (5) Infrared imaging camera, used to monitor the enhanced heat transfer effect of LMMs. Model: 890, Testo, Germany. (6) Field emission environmental scanning electron microscopes, used for characterization of graphite and platinum obstacle surfaces after operation morphology. Model: FESEM, QUANTA FEG 250, America.

## Acknowledgements

This work was partially supported by the National Natural Science Foundation of China under Grants No. 12402324 and No. 91748206.

## Author contributions

JY.L.: methodology, investigation, data curation, formal analysis, validation, visualization, writing- original draft, writing- review and editing; M.G.: visualization; J. W.: writing-review and editing; X. Z.: visualization, writing- review and editing, funding acquisition; J.L.: conceptualization, visualization, supervision, writing- review and editing, funding acquisition.
All authors read and approved this manuscript.

## Conflict of Interest statement

The authors declare no conflict of interest.

## Data Availability statement

The data that support the findings of this study are available from the corresponding author upon reasonable request.